\documentclass[sigconf]{acmart}
\usepackage{multirow}
\AtBeginDocument{%
  }

\copyrightyear{2026}
\acmYear{2026}
\setcopyright{cc}
\setcctype{by}
\acmConference[MM '26]{Proceedings of the 34th ACM International Conference on Multimedia}{November 10--14, 2026}{Rio de Janeiro, Brazil}
\acmBooktitle{Proceedings of the 34th ACM International Conference on Multimedia (MM '26), November 10--14, 2026, Rio de Janeiro, Brazil}
\acmDOI{10.1145/3767308.3835531}
\acmISBN{979-8-4007-2213-4/2026/11}

\begin{document}

\title[PrimitiveCAD: An LLM-Based Point-to-CAD Reconstruction with Primitive-Aware \\ Tokenization and Operation Alignment]{
PrimitiveCAD: An LLM-Based Point-to-CAD Reconstruction with Primitive-Aware Tokenization and Operation Alignment
}


\author{Jian Gao}
\orcid{0009-0008-0391-869X}
\email{244050088@hdu.edu.cn}
\affiliation{%
  \institution{Hangzhou Dianzi University}
  \city{Hangzhou}
  \state{Zhejiang}
  \country{China}
}

\author{Kailin Bi}
\orcid{0009-0002-3342-0282}
\email{242050150@hdu.edu.cn}
\affiliation{%
  \institution{Hangzhou Dianzi University}
  \city{Hangzhou}
  \state{Zhejiang}
  \country{China}
}

\author{Jiamin Xu}
\orcid{0000-0002-7750-7761}
\authornote{Corresponding authors.}
\email{superxjm@yeah.net}
\affiliation{%
  \institution{Hangzhou Dianzi University}
  \city{Hangzhou}
  \state{Zhejiang}
  \country{China}
}

\author{Jinlan Xu}
\orcid{0009-0005-5938-8710}
\email{jlxu@hdu.edu.cn}
\affiliation{%
  \institution{Hangzhou Dianzi University}
  \city{Hangzhou}
  \state{Zhejiang}
  \country{China}
}

\author{Gang Xu}
\orcid{0000-0003-3557-9529}
\authornotemark[1]
\email{gxu@hdu.edu.cn}
\affiliation{%
  \institution{Hangzhou Dianzi University}
  \city{Hangzhou}
  \state{Zhejiang}
  \country{China}
}

\renewcommand{\shortauthors}{Jian Gao, Kailin Bi, Jiamin Xu, Jinlan Xu, \& Gang Xu}

\begin{abstract}

Large-model-based point-to-CAD generation holds immense potential for advancing industrial design and enhancing 3D modeling efficiency. However, most existing methods approach the problem as a general point-cloud encoding and token prediction task, neglecting the tokenization and supervision specifically for CAD-related primitives. As a result, these methods often struggle to accurately reconstruct the intricate primitive structures. To address this limitation, we propose PrimitiveCAD, a novel multi-stage paradigm for point-to-CAD reconstruction that enhances the geometric accuracy of generated CAD models while better preserving critical geometric features. First, we introduce a primitive-aware point cloud tokenization model, enabling the system to learn more robust geometric representations from CAD point clouds. Next, we perform supervised finetuning on a large language model (LLM) and introduce an operation alignment loss to align key CAD operation frequencies, thereby improving the preservation of global shape features. Finally, we incorporate reinforcement learning (RL) and introduce a feature-line alignment reward to further reduce stochasticity and enhance the fine-grained preservation of geometric features. Experiments on the DeepCAD and Fusion360 datasets show that our method achieves state-of-the-art performance in code validity, geometric accuracy, and geometric feature preservation.

\end{abstract}


\begin{teaserfigure}
  \includegraphics[width=\columnwidth]{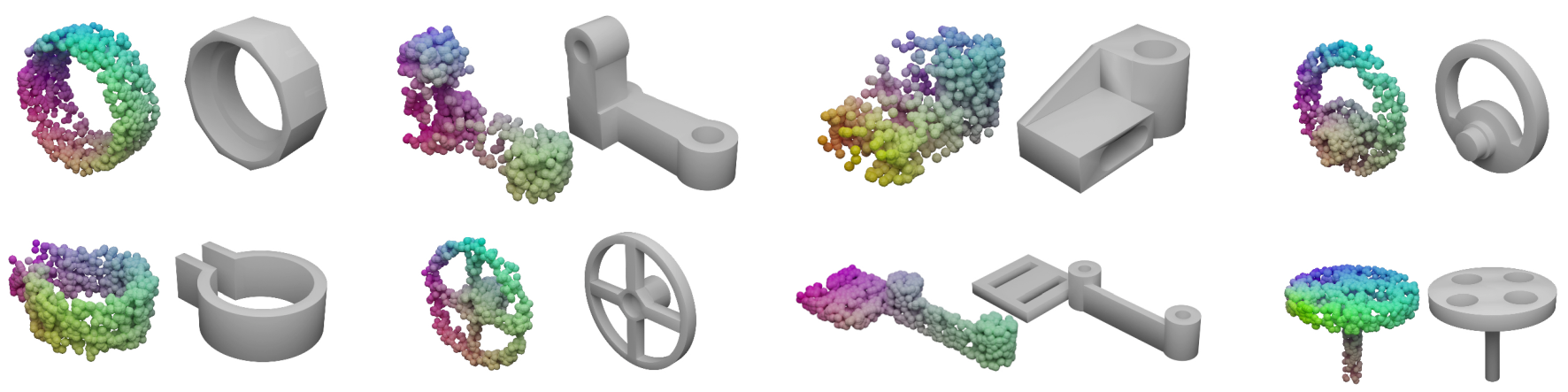}
  \caption{Using a point cloud as input, our method generates an executable Python script that creates a parametric CAD model. Through primitive-aware tokenization and operation-aligned LLM fine-tuning, our approach is capable of producing high-fidelity, feature-preserving CAD models.}
  \label{fig:teaser}
\end{teaserfigure}


\maketitle

\section{Introduction}

Computer-Aided Design (CAD) plays a crucial role in industrial design and manufacturing, enabling the creation and modification of objects across various industries, including architecture, automotive, and aerospace. Traditionally, CAD requires designing 2D sketches and performing 3D operations to generate models—a process that demands specialized knowledge and skills, making it less accessible to non-experts. With the rise of Artificial Intelligence (AI) and Large Language Models (LLMs), a new approach called Computer-Automated Design (CAutoD) has emerged, integrating AI to enhance design speed, creativity, and personalization. This evolution simplifies the CAD generation and manipulation process, making it more intuitive and user-friendly for both experts and non-experts.

For CAD generation, an effective output model representation is essential. Early methods typically relied on Constructive Solid Geometry (CSG)~\cite{foley1996computer} or Boundary Representation (B-Rep)~\cite{ansaldi1985geometric}. Recent studies, however, have shifted towards encoding CAD models as operation sequences~\cite{wu2021deepcad} or more flexible representations, such as Python code (e.g., CadQuery~\cite{rukhovich2025cad}), which offer enhanced interpretability. The task of converting inputs into Python code aligns well with recent advancements in Large Language Models (LLMs), which excel at in-context learning and multimodal tokenization. At the input level, however, due to the limitations of text descriptions and occlusions in 2D images, LLMs often struggle to generate accurate CAD models from simple language or visual descriptions. In contrast, point clouds capture detailed geometric data and can be easily obtained through scanning or direct modeling techniques, making them a superior modality for conditioning and providing more precise guidance for CAD generation. As a result, we condition CAD generation on point cloud inputs and generate CAD models as executable Python scripts.

The point-to-CAD sequence generation method using LLMs converts point cloud data into tokens, which are fed into the LLM as context to generate corresponding CAD sequences using the model's next-token prediction capability. However, current methods have several key limitations: 1) Insufficient capability to extract CAD features from point clouds; existing methods either train point feature extraction from scratch via end-to-end training or use general point cloud feature extractors like PointNet~\cite{qi2017pointnet} or Point Transformer~\cite{zhao2021point,wu2024point}, which are not specifically designed for CAD features; 2) The supervision for supervised finetuning (SFT) of CAD sequences in LLMs is adapted from general chat models, without adjustments for CAD sequences, leading to suboptimal SFT performance; 3) The reward design in Reinforcement Learning (RL) only considers geometric overlap, neglecting the specific point-line-plane structural features inherent in CAD models.

To resolve these issues, our method first trains a primitive-aware tokenizer for point clouds, using Point Transformer v3 (PTv3)~\cite{wu2024point} as the backbone. We fine-tune PTv3 with geometry primitives derived from CAD sketching and extrusion operations, enabling it to produce discriminative features for points sampled from CAD models. Next, we perform SFT to allow the LLM to generate Python code. To enhance SFT supervision, we introduce an operation alignment loss specifically designed for CAD sequences, which aligns the expected frequency of core CAD operations and imposes macro-level geometric constraints. We found that even with SFT alone, our method performs on par with existing baselines that incorporate RL. Finally, similar to Cadrille~\cite{kolodiazhnyi2025cadrille}, we apply RL to further refine the model. In the RL phase, traditional volume-based IoU rewards often fail to capture fine geometric details and provide insufficient guidance when small features are missing. To address this, we propose a geometry feature-line alignment reward that explicitly extracts and matches key topological features. Incorporating this RL phase enables our method to generate CAD models with low stochasticity and highly accurate reconstruction performance. 

In summary, this work systematically integrates CAD structural information into the entire learning pipeline, spanning point-cloud tokenization, SFT, and RL. Our main contributions are summarized as follows:
\begin{itemize}
    \item We propose a point-to-CAD reconstruction framework with a multi-stage training paradigm. This framework first extracts discriminative features to achieve primitive-aware tokenization for CAD point clouds, and then uses an LLM to generate CAD sequences through SFT and RL.
    \item During the SFT phase, we introduce an operation alignment loss that uses a relaxed IoU on the frequency of key operators, establishing macro-level constraints for CAD operation sequences and improving SFT performance.
    \item During the RL phase, we introduce a novel feature-line alignment reward, effectively addressing the limitations of traditional volume-based IoU rewards when handling small geometric features and complex CAD topologies.
\end{itemize}

Experimental results demonstrate that our point-to-CAD reconstruction framework achieves state-of-the-art performance in CAD sequence reconstruction across multiple public benchmarks.

\section{Related work}



\begin{figure*}[!t]
    \centering
    \includegraphics[width=1.0\linewidth]{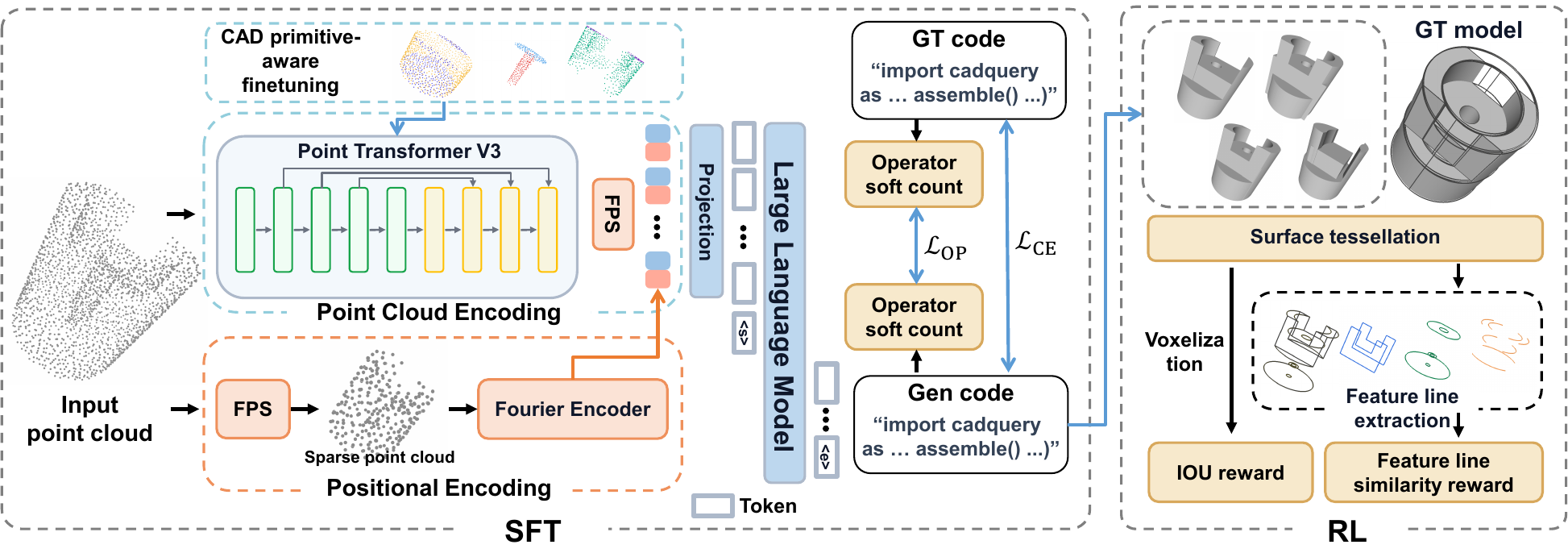}
    \caption{
    Overview of the proposed point-to-CAD reconstruction pipeline. We propose a multi-stage framework for point-to-CAD reconstruction that enhances geometric accuracy while more effectively preserving CAD-specific geometric features. In the first stage, we introduce primitive-aware finetuning to improve point cloud tokenization, allowing the system to learn primitive from CAD point clouds, which are then used as context for the LLM. In the second stage, we fix the tokenization and apply SFT to the LLM, incorporating a novel operation alignment loss to align key CAD operation frequencies, thereby improving the preservation of global shape features. In the third stage, we integrate RL and a feature-line alignment reward to reduce stochasticity and further enhance the fine-grained preservation of geometric features.}
    \label{fig:pipeline}
\end{figure*}

\paragraph{CAD sequence generation}

DeepCAD~\cite{wu2021deepcad} pioneered sequence-based CAD generation by encoding sketch and extrusion operations as discrete tokens. Although CAD-Diffuser~\cite{ma2024draw} and SECAD-Net~\cite{li2023secad} further extended this sequence modeling paradigm, they still rely on customized command spaces or task-specific symbolic sequences. CAD-Recode~\cite{rukhovich2025cad} represents CAD models as Python code based on CadQuery~\cite{developers2024cadquery}, enabling the model to directly generate executable parametric modeling programs rather than customized discrete instructions.

Recent advances in LLMs have revolutionized CAD sequence generation. LLM-based approaches generate CAD construction sequences from text~\cite{khan2024text2cad} or images~\cite{you2025img2cad}. CAD-MLLM~\cite{xu2024cad} further supports text, image, and point-cloud inputs using pretrained multimodal models. CAD-Recode~\cite{rukhovich2025cad} generates Python programs for point-to-CAD reconstruction, while Cadrille~\cite{kolodiazhnyi2025cadrille} extends this paradigm to multimodal inputs. Other approaches improve topology selection~\cite{qi2026pointer}, dimensional precision~\cite{qi2026pointerv2}, and CAD program refinement through visual or geometric feedback~\cite{li2026seek,fan2026caddesigner,kabisov2026cadreasoner}. However, these methods rely solely on token-level cross-entropy loss, which does not account for the unique characteristics of CAD operations, limiting the model's ability to preserve CAD features.

Since large-model generation is inherently stochastic, recent approaches like CADCrafter~\cite{chen2025cadcrafter}, Cadrille~\cite{kolodiazhnyi2025cadrille} and ReCAD~\cite{li2026recad} use reinforcement learning to reduce stochasticity and improve output quality. However, these methods do not explicitly align fine-grained CAD geometric features. To address this, we introduce a feature-line alignment reward that aligns key geometric features (such as lines, arcs, and circles), thereby enhancing local geometric fidelity during reinforcement learning. 

\paragraph{Point-based CAD generation}

Early point-to-CAD methods used parametric surface fitting~\cite{benkHo2004segmentation}, RANSAC~\cite{li2011globfit}, or region growing~\cite{oesau2016planar} for primitive fitting and object decomposition. Later approaches directly recover geometric structures from point clouds but often rely on predefined primitives or local fitting objectives~\cite{liu2024point2cad,wang2020pie,zhu2023nerve}. LLM-based methods generate CAD programs from point clouds using learned point-cloud representations~\cite{xu2024pointllm,liao2025automated,rukhovich2025cad}. However, they lack primitive-aware geometric priors, which may limit representation and reconstruction accuracy. To overcome this limitation, we introduce primitive-aware point cloud tokenization to enhance geometric representation.

\section{Proposed Method}

\subsection{Overview}

Our goal is to generate a CAD model from an input point cloud~$\mathcal{P}$, where the target CAD model is represented as an executable Python script~\cite{rukhovich2025cad} that generates the parametric boundary representation (B-Rep). Consequently, the task of generating a CAD model from a point cloud is transformed into a conditional language generation problem. To achieve this, as shown in Fig.~\ref{fig:pipeline}, we first train a primitive-aware tokenizer for the CAD point cloud (Sec.~\ref{sec:Tokenization}), incorporating point cloud tokens into an LLM~\cite{wang2024qwen2} and then fine-tune it with ground-truth Python scripts. We also introduce an operation alignment loss to enhance supervision (Sec.~\ref{sec:SFT}). However, with SFT, the LLM may still have a probabilistic chance of generating undesirable models. To reduce stochasticity, we incorporate RL and use a novel geometry feature alignment reward to further refine the model (Sec.~\ref{sec:RL}).

\subsection{Primitive-Aware Point Cloud Tokenization for CAD Models}
\label{sec:Tokenization}

We first train a point cloud tokenizer for CAD models, which extracts discriminative features from the point cloud, followed by sampling and projection to generate point cloud tokens as input for the LLM.

\begin{figure}[t]
    \centering    \includegraphics[width=0.9\columnwidth]{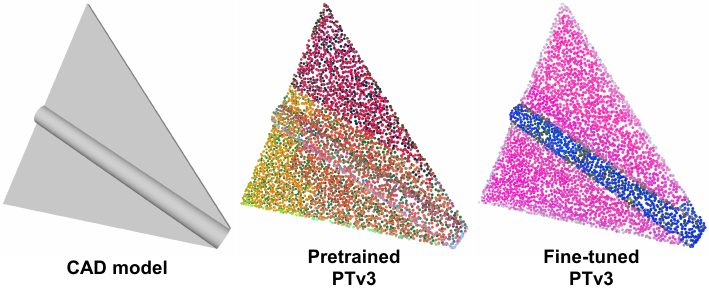}
    \caption{PCA visualization of PTv3~\cite{wu2024point} feature embeddings before and after CAD-specific finetuning. After finetuning with our CAD primitive supervision, the resulting feature space shows enhanced semantic separability.}
    \label{fig:PTv3_feature}
\end{figure}

\begin{figure}[t]
    \centering
    \includegraphics[width=0.9\columnwidth]{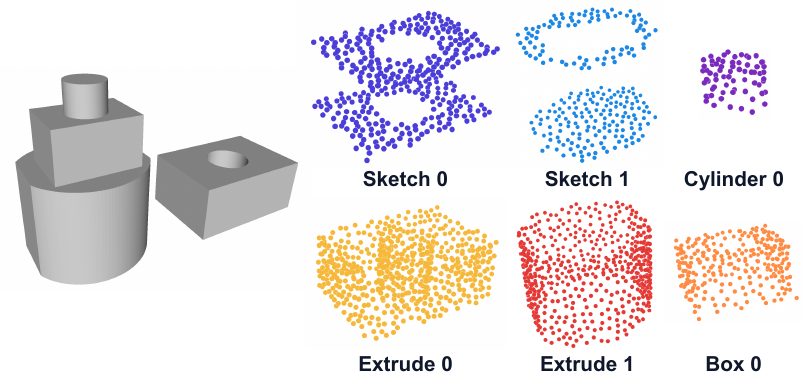}
    \caption{Point-wise primitive instance labels for primitive-aware finetuning of PTv3~\cite{wu2024point}, including instance assignments for sketch, extrude, cylinder, and box primitives.}
    \label{fig:segmentation}
\end{figure}

Regarding point cloud features, the current self-supervised pre-trained Point Transformer v3 (PTv3)~\cite{wu2024point} is designed for 3D point cloud downstream tasks, such as indoor and outdoor scene segmentation. However, as shown in Fig.~\ref{fig:PTv3_feature}, directly applying these pre-trained features is not suitable for CAD models, which involve operations like sketching, extrusion, and boolean operations. To achieve more descriptive tokenization, we fine-tune the pre-trained PTv3 network by adding two heads, enabling it to output two types of predictions for each point $i$: a normal vector $\hat{\mathbf{n}}_i \in \mathbb{R}^{3}$ and a 32-dimensional category logit vector $\hat{\mathbf{s}}_i \in \mathbb{R}^{32}$, where $\hat{\mathbf{s}}_i$ is a probability distribution for CAD-primitive instance segmentation: 
\begin{equation}
\hat{\mathbf{S}} = \mathrm{softmax}\left(\begin{bmatrix} \hat{\mathbf{s}}_1^\top \\ \vdots \\ \hat{\mathbf{s}}_N^\top \end{bmatrix}\right) \in [0,1]^{N \times 32}.
\end{equation}
Here, $N$ is the number of points. Each $\hat{\mathbf{s}}_i$ represents a fine-grained segmentation of the point cloud into non-overlapping primitive instances. The primitive instances are categorized into four types: 1) sketch, 2) extrude, 3) cylinder, and 4) box, as shown in Fig.~\ref{fig:segmentation}. Each type contains at most $K=8$ instances.

\paragraph{Ground-Truth Primitive Labels}

Given the CAD sequence and the sampled point cloud, we can obtain the ground-truth primitive labels $\mathbf{S} \in [0,1]^{N \times 32}$ by parsing the geometric instances and extracting the necessary parameters for label assignment. Each point's primitive-instance label is determined by comparing its position to the geometric constraints of the parsed instances. If a point lies within the support region of an instance, it is assigned to that instance. For sketch and extrude primitives, we check whether the point lies on the base or top plane for the base-face label and on the barrel surface for the barrel-face label. For cylinder primitives, we verify if the point belongs to the cylindrical surface based on its radial and axial position. For box primitives, label assignment is based on the local coordinate system. This process ensures that points are consistently mapped to the correct primitive-level instance labels based on geometric constraints.


\paragraph{Primitive Instance Segmentation Supervision}

After obtaining the ground-truth primitive labels for the point cloud, we fine-tune the PTv3~\cite{wu2024point}. The training objective includes an unoriented normal consistency loss, $\mathcal{L}_\mathrm{n}$, and a primitive instance segmentation loss, $\mathcal{L}_\mathrm{seg}$, which together constrain the per-point geometric direction and the semantic classification of points to primitive instances, such as base-or-barrel for sketch/extrude operations, and cylinder or box for other primitives. The overall loss function is expressed as:
\begin{equation}
\mathcal{L}_\mathrm{pt} = \mathcal{L}_\mathrm{n} + \lambda_\mathrm{pt} \mathcal{L}_\mathrm{seg}.
\end{equation}

To resolve the ambiguity caused by normal sign flipping, we constrain the predicted and ground-truth normals using an unoriented cosine distance:
\begin{equation}
\mathcal{L}_\mathrm{n} = 1 - \frac{1}{N} \sum_{i=1}^N | \hat{\mathbf{n}}_{i} \cdot \mathbf{n}_i |.
\end{equation}




\paragraph{Primitive Instance Segmentation Loss}

Since the indices of primitive instances are not ordered, different permutations of the indices may correspond to the same geometric partition. Thus, we treat the predictions as an unordered set and use Hungarian matching~\cite{kuhn1955hungarian} to find the optimal one-to-one correspondence with the ground truth primitive labels. We construct a $32 \times 32$ similarity matrix and apply Hungarian matching~\cite{kuhn1955hungarian} to obtain the permutation $\pi_g$, which reorders the predicted instances into $\pi(\hat{\mathbf{S}})$ to align with $\mathbf{S}$.

Next, we compute the instance segmentation loss $\mathcal{L}_{seg}$ by applying the Relaxed IoU (RIoU) similarity between the predicted instances and the ground-truth instances:
\begin{align}
\mathcal{L}_\mathrm{seg} &= 1 - \text{RIoU}(\pi(\hat{\mathbf{S}}), \mathbf{S}),\\
\text{RIoU}(X, Y) &= \frac{X^\top Y}{\|X\|_1 + \|Y\|_1 - X^\top Y}.
\end{align}

Through finetuning on normal estimation and primitive instance segmentation tasks, the PTv3 model~\cite{wu2024point} gains a clear understanding of the geometric semantics and spatial structure of CAD instances (Fig.~\ref{fig:segmentation}). This backbone network is then used as a geometric feature extractor, transforming the point cloud into the latent features $\mathbf{F} \in \mathbb{R}^{N \times 64}$ which provides geometric priors and primitive parsing capabilities for subsequent CAD code generation.

\paragraph{Latent Features Tokenization}

To transform the latent features $\mathbf{F} \in \mathbb{R}^{N \times 64}$ into tokens $\mathbf{T} \in \mathbb{R}^{N \times D}$, where $D$ represents the token dimension, we first apply farthest point sampling (FPS) to select a subset of 256 points. We then extract features from the full set of points and obtain the features $\mathbf{F}_{256} \in \mathbb{R}^{256 \times 64}$ for the selected subset. To mitigate the loss of continuous spatial values, the encoder applies a Fourier transform to the original coordinates of $\mathbf{P}_{256} \in \mathbb{R}^{256 \times 3}$, generating high-frequency positional encodings that capture absolute spatial coordinate information:
\begin{equation}
\phi(x) = [x, \sin(\omega_1 x), \cos(\omega_1 x), \dots, \sin(\omega_M x), \cos(\omega_M x)],
\end{equation}
where $\omega_m$ denotes a predefined set of frequencies. We then stack the latent features $\mathbf{F}_{256}$ and the positional encodings $\phi(\mathbf{P}_{256})$, and tokenize them through a linear projection, yielding the final point cloud token sequence:
\begin{equation}
\mathbf{T}_{pc} = \text{Linear}\left([\mathbf{F}_{256},\phi(\mathbf{P}_{256})]\right) \in \mathbb{R}^{256 \times D}.
\end{equation}

\subsection{SFT with Operation Alignment}
\label{sec:SFT}

To achieve deep alignment between geometric information and textual semantics in a unified representation space, we fine-tune the Qwen2 LLM~\cite{wang2024qwen2} through SFT. We reserve a placeholder interval of length 256 in the input sequence of the language model and replace the corresponding token embeddings within this interval with point cloud tokens $\mathbf{T}_{pc}$. This explicit feature alignment strategy allows the model to directly map geometric structure priors and coordinate representations into the reasoning space of the LLM, providing robust cross-modal contextual support for subsequent CAD code generation.

To better supervise the model, we decouple the training objective into two components: a traditional token-level autoregressive generation loss and a novel operation-level global alignment loss specifically designed for CAD sequences. The former ensures the local correctness of the code syntax, while the latter enhances the model's ability to capture the global geometric characteristics by aligning the frequency of core CAD operations.

\begin{figure}[t]
    \centering
    \includegraphics[width=\columnwidth]{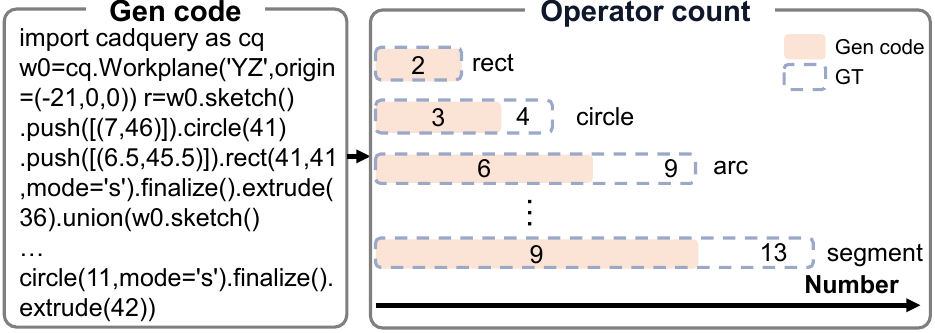}
    \caption{Comparison of operator counts between the generated CAD sequence and the ground truth. Differences across operator types reflect the operation-level mismatch targeted by the operation alignment loss.}
    \label{fig:operator_count}
\end{figure}

\paragraph{Autoregressive Cross-Entropy Loss}

In line with the standard finetuning paradigm for large language models, we compute the negative log-likelihood of the target code sequence as the primary sequence modeling loss. Given the target CAD code sequence $C = \{c_1, c_2, \dots, c_n\}$, the cross-entropy loss $\mathcal{L}_\mathrm{ce}$ is defined as:
\begin{equation}
\mathcal{L}_\mathrm{ce} = -\frac{1}{n} \sum_{i=1}^n \log P(c_i \mid c_{<i}, \mathbf{T}_{pc}).
\end{equation}
This loss guides the model in generating syntactically valid Python code from the extracted point cloud tokens $\mathbf{T}_{pc}$.

\subsubsection{Operation Alignment Loss}

Although $\mathcal{L}_\mathrm{ce}$ effectively guides token-level code generation, CAD reconstruction is highly sensitive to operation-level errors. Missing or redundant operators can lead to significant deviations in the final geometry. To address this, we introduce an operation alignment loss $\mathcal{L}_\mathrm{op}$ that regularizes the global operator composition of the generated CAD sequence.

Let $\mathcal{O}$ be the set of core CAD operators and primitive types. For each ground-truth Python code, we calculate the occurrence count $C_o^{gt}$ for each operator $o \in \mathcal{O}$ using explicit string matching. In contrast, for the predicted Python code, directly counting operator occurrences from discrete tokens is non-differentiable. To address this, we define a differentiable soft count $C_o^{pr}$ by summing the operator's probabilities. Since each operator can have multiple token patterns, we enumerate the valid tokenization variants. We calculate the matching probability for each pattern at every position in the sequence, aggregate the variants in log space, and sum the probabilities across all positions to obtain the soft count $C_o^{pr}$. As shown in Fig.~\ref{fig:operator_count}, discrepancies in operator counts indicate global inconsistencies in the CAD construction logic.

Based on the ground-truth count vector and the predicted soft count vector, the operation alignment loss is defined as:
\begin{equation}
\mathcal{L}_\mathrm{op} = 1 - \frac{1}{B} \sum_{b=1}^{B} \mathrm{RIoU}(C_o^{gt}, C_o^{pr}).
\end{equation}
The overall training objective combines the token-level autoregressive loss and the operation alignment loss as follows:
\begin{equation}
\mathcal{L}_\mathrm{sft} = \mathcal{L}_\mathrm{ce} + \lambda_\mathrm{sft} \mathcal{L}_\mathrm{op}.
\end{equation}

By introducing $\mathcal{L}_\mathrm{op}$ as an auxiliary regularization term, the model is guided not only to generate syntactically valid CAD code, but also to maintain the underlying construction logic and structural characteristics of the target CAD geometry.

\begin{figure}[t]
    \centering
    \includegraphics[width=0.9\columnwidth]{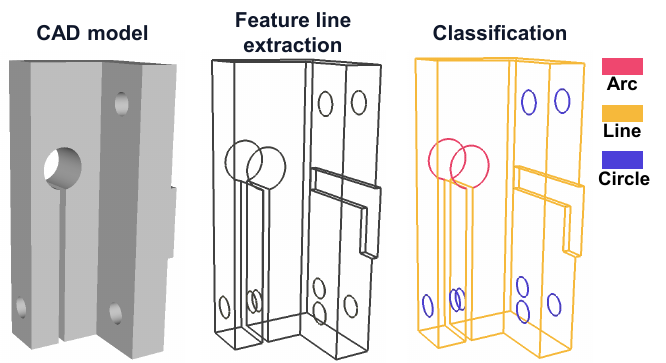}
    \caption{Feature line extraction and classification in the RL stage. The extracted features are categorized into three geometric types: straight lines, arcs, and circles.}
    \label{fig:feature_line}
\end{figure}

\subsection{RL with Feature-Line Alignment}
\label{sec:RL}

After the SFT stage, we introduce an online reinforcement learning phase where the model is further refined using feedback from programmatic execution to reduce stochasticity and improve geometric accuracy. We use Group Relative Policy Optimization (GRPO)~\cite{shao2024deepseekmath} for feedback, which does not require a separate reward model. Instead, it estimates an advantage function based on the relative quality of different samples within a group.

At this stage, the paired point cloud and code sequence used for SFT is no longer required. Unlike SFT which relies on annotated CAD reconstruction sequences, RL evaluates the results of the model-generated outputs through execution and feeds the reward signals back into the optimization process. This eliminates the need for complete CAD sequence labels and allows the use of 3D datasets without sequence annotations, significantly expanding the available training data. By learning from more diverse geometric shapes, the LLM becomes more robust to geometric noise and artifacts in real-world scenarios.

In each training iteration, given the point cloud tokens, the fine-tuned LLM samples $K$ candidate code sequences ${C_1, C_2, \dots, C_K}$ in parallel. We execute the generated code to obtain the corresponding CAD models and compute the reward $r$ for each sample. The advantage function $A$ is calculated based on the normalized rewards within the group. To evaluate the geometric quality of the generated sequences, we design a reward function that considers both global volumetric consistency and local feature fidelity, addressing the limitations of conventional volumetric rewards in capturing fine-grained structures.

When the generated code successfully executes into a valid CAD model $M^{pred}$, we align and normalize both the predicted model and the ground-truth model $M^{gt}$. The global geometric similarity is then measured using the IoU between $M^{gt}$ and $M^{pred}$. Additionally, as shown in Fig.~\ref{fig:feature_line}, local structural similarity $S^{feat}$, captured through feature-line extraction and matching, is incorporated into the reward function:
\begin{equation}
r = w_\mathrm{iou} \cdot \mathrm{IoU}(M^{gt}, M^{pred}) + w_\mathrm{feat} \cdot S^{feat}(M^{gt}, M^{pred}).
\end{equation}
We set $w_\mathrm{iou}=0.6$ and $w_\mathrm{feat}=0.4$. When the generated code fails to execute, we fix the reward $r$ to a value of $-10$.

To compute $S^{feat}$, we perform feature-line matching by calculating the similarity for each feature type (e.g., straight lines, arcs, and circles). This is done by constructing a normalized similarity matrix based on the $1 - L2$ distance, which incorporates geometric attributes such as endpoint distance, center offset, length ratio, and directional consistency. Since the numbers of feature lines in the ground-truth and predicted models may differ, $S_{\mathrm{feat}}$ is computed as the optimal normalized matching score using the Hungarian algorithm~\cite{kuhn1955hungarian}, where unmatched feature lines are assigned zero score in the final aggregation. For feature-line extraction, we extract feature lines in the mesh domain; please refer to the supplementary materials for further information.

\section{Experiments}

To validate the effectiveness of the proposed method, we compare it with representative methods for point-to-CAD reconstruction. The baselines include sketch and extrusion sequence modeling methods, such as DeepCAD~\cite{wu2021deepcad}, Point2Cyl~\cite{uy2022point2cyl}, HNC-CAD~\cite{xu2023hierarchical}, MultiCAD~\cite{ma2023multicad}, TransCAD~\cite{dupont2024transcad}, CAD-Diffuser~\cite{ma2024draw}, and CAD-SIGNet~\cite{khan2024cad}, as well as recent methods that generate executable programs, including CAD-Recode~\cite{rukhovich2025cad} and Cadrille~\cite{kolodiazhnyi2025cadrille}. In the ablation study, we examine the effectiveness of the PTv3~\cite{wu2024point} finetuning and the operation alignment during the SFT stage. Then we compare the reconstruction performance achieved without feature-line alignment reward and with different RL strategies. We also assess the impact of point cloud tokenization under varying input point counts. For implementation details, please refer to the supp. material.

\paragraph{Datasets} During the SFT stage, to ensure a fair comparison with existing methods for point-to-CAD reconstruction, we adopt the DeepCAD dataset~\cite{wu2021deepcad}, which contains approximately 16k CAD models collected from real design data, as the primary training set for point cloud based supervision. We also incorporate approximately one million models from the CAD-Recode dataset~\cite{rukhovich2025cad} for training, where all CAD models are constructed from executable programs written in CadQuery~\cite{developers2024cadquery}. During the RL stage, we leverage the Fusion360 dataset~\cite{willis2021fusion}, a small scale CAD reconstruction benchmark containing complex and realistic CAD models, to guide the model toward generating CAD structures that better satisfy real world engineering requirements.

\begin{table*}[h]
  \centering
  \small
  \caption{Comparison of point-to-CAD methods on the DeepCAD and Fusion360 datasets~\cite{wu2021deepcad,willis2021fusion}, training on DeepCAD~\cite{wu2021deepcad} (DCAD), CAD-Recode~\cite{rukhovich2025cad} (CADR), or both datasets. Our method consistently outperforms existing state of the art methods. For fair comparison, all reported results are obtained without RL finetuning or test-time sampling.}
  \label{tab1}
  \begin{tabular*}{\textwidth}{@{\extracolsep{\fill}} lll cccc cccc @{}}
    \toprule
    \multirow{2}{*}[-2pt]{Method} & \multirow{2}{*}[-2pt]{Year} & \multirow{2}{*}[-2pt]{Train Dataset} 
    & \multicolumn{4}{c}{DeepCAD Test Set} 
    & \multicolumn{4}{c}{Fusion360 Test Set} \\
    \cmidrule(lr){4-7} \cmidrule(lr){8-11}
    & 
    &
    & Mean CD$\downarrow$ & Med.\ CD$\downarrow$ & IoU$\uparrow$ & IR$\downarrow$
    & Mean CD$\downarrow$ & Med.\ CD$\downarrow$ & IoU$\uparrow$ & IR$\downarrow$ \\
    \midrule
    DeepCAD
    & 2021   & DCAD    & 42.50    & 9.64  & 46.7 & 7.1  & 76.10    & 89.20 & 39.9 & 25.2 \\
    Point2Cyl   & 2022 & DCAD    & --    & 4.27  & 73.8 & 3.9  & --    & 4.18 & 67.5 & 3.2  \\
    HNC-CAD  & 2023    & DCAD    & --    & 8.64  & 65.3 & 5.6  & --    & 36.80 & 63.5 & 7.3  \\
    MultiCAD  &2023    & DCAD    & --    & 8.09  & --   & 11.5 & --    & 42.20 & --   & 16.5 \\
    TransCAD  &2024   & DCAD    & 32.30    & 4.51  & 65.5 & 1.1  & 78.60    & 33.40 & 60.2 & 2.4  \\
    CAD-Diffuser  & 2024 & DCAD    & --    & 3.02  & 74.3 & 1.5  & --    & 3.85 & 63.2 & 1.7  \\
    CAD-SIGNet  & 2024  & DCAD    & 6.81    & 0.29  & 77.3 & 5.0  & 14.50    & 0.70 & 58.3 & 9.3  \\
    Cadrille   & 2025   & DCAD    & 3.43    & 0.25  & 79.4 & 0.4  & 7.61    & 0.66 & 63.7 & 0.6  \\
    Ours & 2026 & DCAD & \textbf{0.75} & \textbf{0.18} & \textbf{86.0} & \textbf{0.1} & \textbf{0.81} & \textbf{0.19} & \textbf{82.0} & \textbf{0.5} \\
    \midrule
    CAD-Recode   & 2025  & CADR & 0.83    & 0.18  & 87.1 & 3.1  & 1.21    & 0.19 & 79.1 & 5.0  \\
    Cadrille  & 2025   & CADR & 0.76    & 0.18  & 87.1 & 2.1  & 1.10    & 0.19 & 79.8 & 2.8  \\
    Ours & 2026 & CADR & \textbf{0.69} & \textbf{0.17} & \textbf{89.4} & \textbf{1.9} & \textbf{0.69} & \textbf{0.17} & \textbf{81.7} & \textbf{1.8} \\
    \midrule
    Cadrille  & 2025   & DCAD + CADR & 1.37    & 0.19  & 86.6 & 0.9  & 1.05    & 0.22  & 76.5 & 2.0  \\
    Ours & 2026 & DCAD + CADR & \textbf{0.44} & \textbf{0.17} & \textbf{90.9} & \textbf{0.1} & \textbf{0.65} & \textbf{0.17} & \textbf{84.9} & \textbf{0.4} \\
    \bottomrule
  \end{tabular*}
\end{table*}

\begin{figure*}[!t]
    \centering
    \includegraphics[width=0.95\linewidth]{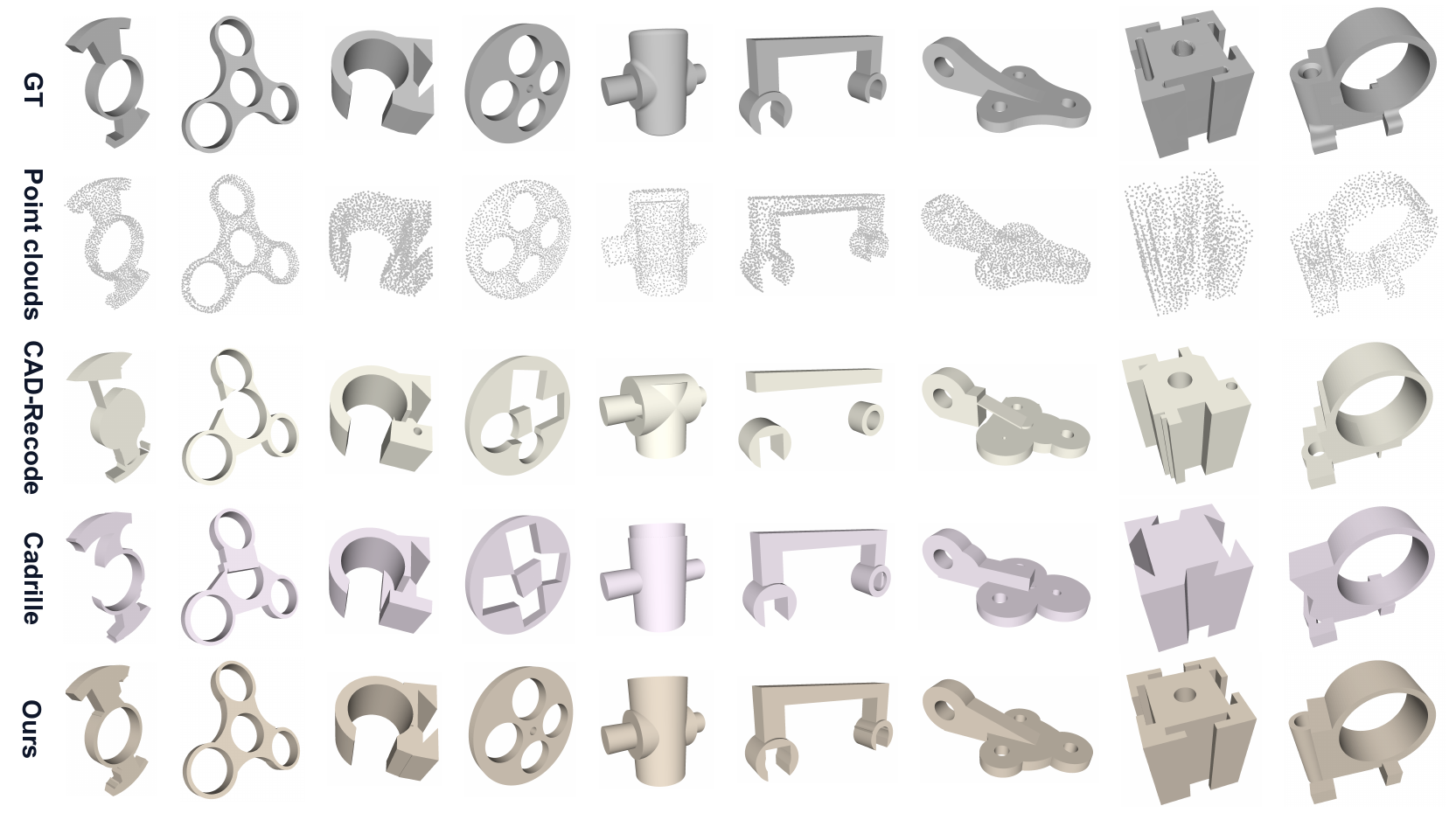}
    \caption{Qualitative CAD reconstruction results. Best result among 10 generated candidates for each method. Compared to prior approaches, the proposed method produces reconstructions with more accurate overall shapes and better-preserved fine-grained geometric features.}
    \label{fig:result}
\end{figure*}

\paragraph{Metrics} Following CAD-Recode~\cite{rukhovich2025cad}, we evaluate our method using four metrics: intersection over union (IoU), invalidity rate (IR), mean Chamfer distance (Mean CD), and median Chamfer distance (Med. CD). IoU measures volumetric overlap with the ground truth, IR is the fraction of code failing to compile into valid 3D solids, and CD assesses surface accuracy by sampling 8192 points per model. Both mean and median CD are scaled by $10^3$ for easier comparison.

For quantitative comparisons, we adopt a single-pass evaluation as in Cadrille~\cite{kolodiazhnyi2025cadrille}, without test-time best-of-$N$ candidate selection. For qualitative comparisons with other LLM-based methods in Fig.~\ref{fig:result}, we report the best result among 10 generated candidates for each method. For the RL ablation in Fig.~\ref{fig:rl_ablation} and Table~\ref{tab5}, we further report both best-of-10 and worst-of-10 results to analyze the effect of RL in reducing stochasticity.


\subsection{Evaluation of SFT and RL}

\begin{table}[t]
  \centering
  \caption{Quantitative comparison of SFT and RL variants on Fusion360~\cite{willis2021fusion}.}
  \label{tab2}
  \begin{tabular*}{\columnwidth}{@{\extracolsep{\fill}} l cccc @{}}
    \toprule
    Method & Mean CD$\downarrow$ & Med.\ CD$\downarrow$ & IoU$\uparrow$ & IR$\downarrow$ \\
    \midrule
    Cadrille w/o RL & 1.05 & 0.220  & 76.5 & 2.00  \\
    Cadrille w/ RL  & 0.58 & 0.176 & 84.8 & 0.46 \\
    Ours w/o RL     & 0.65 & 0.168 & 84.9 & 0.35 \\
    Ours w/ RL & \textbf{0.45} & \textbf{0.161} & \textbf{86.6} & \textbf{0.06} \\
    \bottomrule
  \end{tabular*}
\end{table}

\begin{figure}[t]
    \centering
    \includegraphics[width=\linewidth]{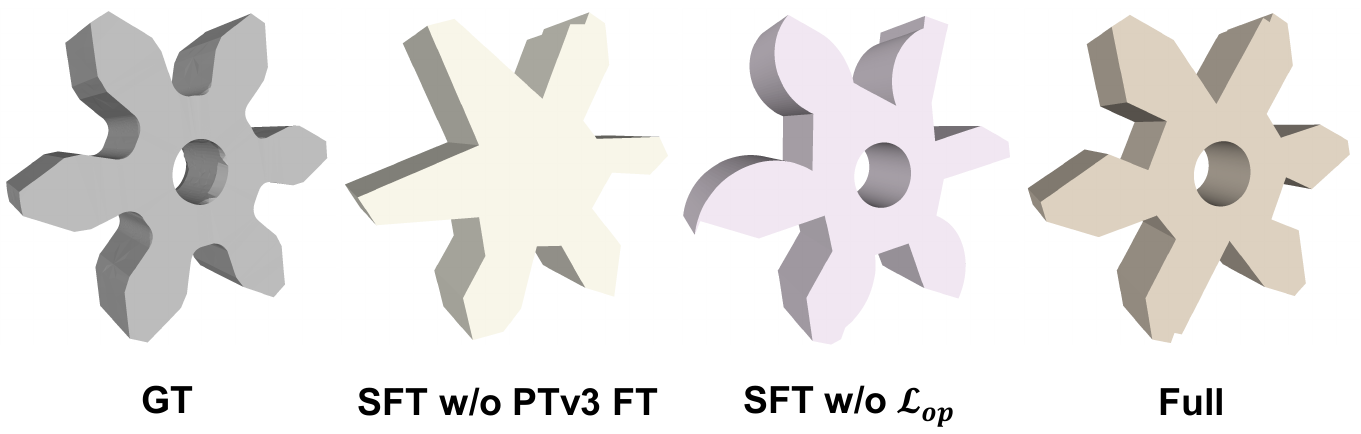}
    \caption{Visualization of the SFT ablation study. We compare variants with and without PTv3~\cite{wu2024point} finetuning and with and without the operation alignment loss $\mathcal{L}_\mathrm{op}$.}
    \label{fig:sft_ablation}
\end{figure}

To comprehensively evaluate the proposed method, we conduct quantitative comparisons with representative prior methods on the DeepCAD~\cite{wu2021deepcad} and Fusion360~\cite{willis2021fusion} test sets, and further compare our method with Cadrille~\cite{kolodiazhnyi2025cadrille} under both the SFT-only and RL-finetuned settings. The quantitative results are reported in Table~\ref{tab1} and Table~\ref{tab2}, while qualitative comparisons are shown in Fig.~\ref{fig:result}.

As shown in Table~\ref{tab1}, our method achieves the best overall performance on both DeepCAD~\cite{wu2021deepcad} and Fusion360~\cite{willis2021fusion}. In particular, under the same Qwen2VL-2B-Direct~\cite{wang2024qwen2} backbone as Cadrille~\cite{kolodiazhnyi2025cadrille}, our method obtains higher IoU and lower Mean CD and median CD, indicating more accurate CAD reconstruction. As shown in the last two rows of Table~\ref{tab1}, the performance of our method can be further enhanced by training on the larger CAD-Recode~\cite{rukhovich2025cad}. These results validate the effectiveness of the primitive-aware tokenization and the proposed SFT design.

In Table~\ref{tab2}, we evaluate the effectiveness of RL by comparing our method with Cadrille~\cite{kolodiazhnyi2025cadrille} under both the SFT-only and RL-finetuned settings on the Fusion360 dataset~\cite{willis2021fusion}. Even without RL, our model already achieves performance on par with the RL-finetuned Cadrille~\cite{kolodiazhnyi2025cadrille}. After RL finetuning, our method consistently outperforms Cadrille~\cite{kolodiazhnyi2025cadrille} on all evaluation metrics, demonstrating that the proposed RL stage further improves reconstruction accuracy and validity on top of the SFT baseline. In Fig.~\ref{fig:result}, we present qualitative CAD reconstruction results from input point clouds. Compared with CAD-Recode~\cite{rukhovich2025cad} and Cadrille~\cite{kolodiazhnyi2025cadrille}, our approach can better-preserved fine-grained geometric features, such as holes and right angles. These qualitative improvements are consistent with the quantitative gains in Table~\ref{tab1} and Table~\ref{tab2}, and can be attributed to the combined effect of PTv3~\cite{wu2024point} finetuning with primitive labels, the operation alignment during the SFT stage, and the feature-line alignment reward adopted during the RL stage. We also compare our method with CAD-MLLM~\cite{xu2024cad}, please refer to supp. material.

We further train and evaluate our method on the complex CAD dataset CADEvolve~\cite{elistratov2026cadevolve}, demonstrating that our framework can generalize to complex CAD operations, including fillet, chamfer, and shell. The results are provided in the supplementary material, demonstrating that our method can reconstruct CAD models with these advanced operations while preserving geometric accuracy.

\subsection{Ablations}

\paragraph{Primitive-Aware Tokenization and Operation Alignment in SFT} To evaluate the effectiveness of PTv3~\cite{wu2024point} finetuning with primitive labels and the proposed operation alignment loss, we conduct an ablation study during the SFT stage. The qualitative and quantitative results are shown in Fig.~\ref{fig:sft_ablation} and Table~\ref{tab3}, respectively. Table~\ref{tab3} shows that PTv3~\cite{wu2024point} finetuning with primitive labels and the introduction of $\mathcal{L}_{op}$ jointly improve IoU and CD, enhancing overall reconstruction performance. The qualitative comparisons in Fig.~\ref{fig:sft_ablation} are consistent with these quantitative gains. For further ablation studies, please refer to the supp. material.

\paragraph{Number of Points for Tokenization} To analyze the impact of input point numbers on tokenization, we evaluate four different input point quantities. Table~\ref{tab4} shows that the variant using 512 input points performs the worst in terms of IoU and CD, indicating that 512 points are insufficient for learning effective primitive-aware geometric representations. The variants using 1024, 2048, and 4096 input points achieve comparable overall performance. In particular, the 1024-point variant obtains the best IoU, while the 4096-point variant achieves slightly better CD. These results indicate that increasing the number of input points beyond 1024 brings only marginal gains. Overall, the results suggest that 1024 input points provide a good balance between reconstruction performance and input complexity, and are sufficient for learning effective geometric representations.

\begin{table}[t]
  \centering
  \caption{Ablation study of the operation alignment loss during supervised finetuning.}
  \label{tab3}
  \begin{tabular*}{\columnwidth}{@{\extracolsep{\fill}} l cccc @{}}
    \toprule
    Method & Mean CD$\downarrow$ & Med.\ CD$\downarrow$ & IoU$\uparrow$ & IR$\downarrow$ \\
    \midrule
    SFT w/o PTv3 FT & 0.88    & 0.177 & 82.4  & 0.70   \\
    SFT w/o $\mathcal{L}_\mathrm{op}$ & 0.71 & 0.172 & 83.4 & 0.75 \\
    SFT & \textbf{0.65} & \textbf{0.168} & \textbf{84.9} & \textbf{0.35} \\
    \bottomrule
  \end{tabular*}
\end{table}

\begin{table}[t]
  \centering
  \caption{Ablation study with varying numbers of input points for primitive-aware point cloud tokenization.}
  \label{tab4}
  \begin{tabular*}{\columnwidth}{@{\extracolsep{\fill}} l cccc @{}}
    \toprule
    Method & Mean CD$\downarrow$ & Med.\ CD$\downarrow$ & IoU$\uparrow$ & IR$\downarrow$ \\
    \midrule
    Ours$_{512}$  & 0.66    & 0.171 & 84.4   & 0.35 \\
    Ours$_{1024}$ & 0.65    & 0.168 & 84.9   & 0.35 \\
    Ours$_{2048}$ & 0.66    & 0.170 & 84.6   & 0.46 \\
    Ours$_{4096}$ & 0.62    & 0.167 & 84.6   & 0.41 \\
    \bottomrule
  \end{tabular*}
\end{table}

\begin{figure}[t]
    \centering
    \includegraphics[width=\linewidth]{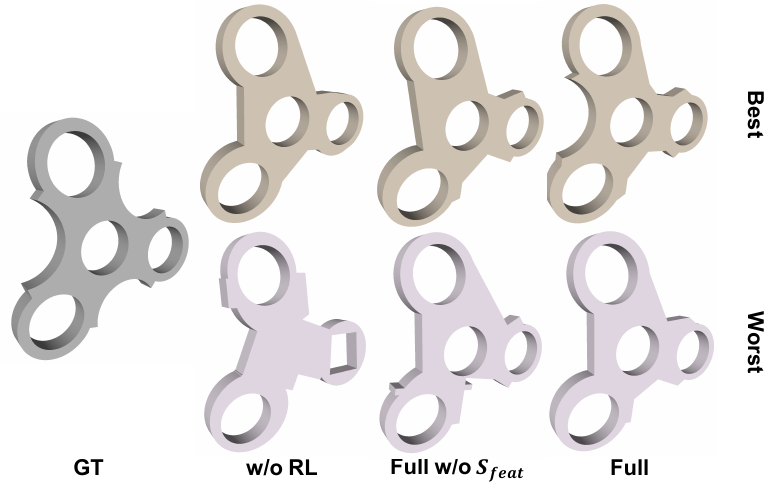}
    \caption{Ablation study on the feature-line alignment reward and RL strategy. Best/worst results over 10 generations are shown.}
    \label{fig:rl_ablation}
\end{figure}

\paragraph{Feature-Line Alignment Reward and RL Strategy} To evaluate the effect of the feature-line alignment reward, we conduct an ablation study with three variants: the SFT-only baseline, an RL model without the feature-line alignment reward, and the full RL model. To demonstrate that RL can effectively reduce stochasticity, we report the best and worst results out of 10 trials. As shown in Table~\ref{tab5} and Fig.~\ref{fig:rl_ablation}, RL improves reconstruction robustness compared to the SFT-only baseline, particularly under worst-of-10 evaluation. Removing the feature-line alignment reward reduces IoU and increases CD, demonstrating its contribution to fine-grained geometric supervision, especially in preserving feature-line boundaries and subtle geometric details. 

We also compare the GRPO~\cite{shao2024deepseekmath} strategy used in our RL phase with the Dr.\ CPPO strategy used in Cadrille~\cite{kolodiazhnyi2025cadrille}. Under the same reward design, Dr.\ CPPO~\cite{kolodiazhnyi2025cadrille} performs worse than the full model. This may be because Dr. CPPO~\cite{kolodiazhnyi2025cadrille} relies solely on the best and worst samples for preference construction, potentially overlooking valuable supervision from the remaining candidates, which can reduce training stability. In contrast, the full model exhibits small best--worst gaps (mean CD gap = 0.59, median CD gap = 0.037, IoU gap = 7.3, IR gap = 0.87), indicating better and more stable performance across sampled candidates.

\begin{table}[t]
  \centering
  \caption{Ablation study on the feature-line alignment reward and RL strategy. Best/worst results over 10 generations are shown.}
  \label{tab5}
  \setlength{\tabcolsep}{2.5pt}
  \begin{tabular*}{\columnwidth}{@{\extracolsep{\fill}} l cccc @{}}
    \toprule
    Method & Mean CD$\downarrow$ & Med.\ CD$\downarrow$ & IoU$\uparrow$ & IR$\downarrow$ \\
    \midrule
    w/o RL                & 0.27/1.41 & 0.143/0.210 & 89.5/76.2 & 0.00/3.88 \\
    Full w/o $S_{feat}$   & 0.28/0.87 & 0.146/0.186 & 89.6/81.5 & 0.00/0.93 \\
    Full (Dr.\ CPPO)      & 0.31/0.97 & 0.149/0.188 & 89.4/81.2 & 0.00/1.10 \\
    Full                  & \textbf{0.26/0.85} & \textbf{0.142/0.179} & \textbf{89.6/82.3} & \textbf{0.00/0.87} \\
    \bottomrule
  \end{tabular*}
\end{table}

\section{Conclusion}


We propose PrimitiveCAD, a multi-stage approach for generating CAD sequences from point clouds. PTv3 fine-tuning with primitive labels improves geometric representations, operation alignment loss enhances global shape accuracy during SFT, and feature-line alignment reward refines geometric details during RL. Experiments demonstrate the effectiveness of our approach in terms of geometric accuracy, code validity, and reconstruction quality.

\paragraph{Limitation} Currently, our framework mainly focuses on component level generation and does not yet support complex multi-module assembly models, which will be explored in future work.


\begin{acks}
This research was supported by the National Key R\&D Program of China
(No. 2023YFB3309103), the ``Pioneer'' and ``Leading Goose'' R\&D Program
of Zhejiang Province (No. 2025C01086), the National Natural Science
Foundation of China (No. 62572161 and 62302134), the Natural Science
Foundation of Zhejiang Province (No. LQ24F020031), and the Hangzhou Key
Scientific Research Program (No. 2025SZD1A48).
\end{acks}





\bibliographystyle{ACM-Reference-Format}
\bibliography{acmart}

@article{ansaldi1985geometric,
  title={Geometric modeling of solid objects by using a face adjacency graph representation},
  author={Ansaldi, Silvia and De Floriani, Leila and Falcidieno, Bianca},
  journal={ACM SIGGRAPH Computer Graphics},
  volume={19},
  number={3},
  pages={131--139},
  year={1985},
  publisher={ACM New York, NY, USA}
}

@article{benkHo2004segmentation,
  title={Segmentation methods for smooth point regions of conventional engineering objects},
  author={Benk{\H{o}}, P{\'a}l and V{\'a}rady, Tam{\'a}s},
  journal={Computer-Aided Design},
  volume={36},
  number={6},
  pages={511--523},
  year={2004},
  publisher={Elsevier}
}

@inproceedings{chen2025cadcrafter,
  title={Cadcrafter: Generating computer-aided design models from unconstrained images},
  author={Chen, Cheng and Wei, Jiacheng and Chen, Tianrun and Zhang, Chi and Yang, Xiaofeng and Zhang, Shangzhan and Yang, Bingchen and Foo, Chuan-Sheng and Lin, Guosheng and Huang, Qixing and others},
  booktitle={Proceedings of the IEEE/CVF Conference on Computer Vision and Pattern Recognition},
  pages={11073--11082},
  year={2025}
}

@misc{developers2024cadquery,
  title={Cadquery: A python parametric cad scripting framework},
  author={Developers, CADQuery},
  year={2024}
}

@inproceedings{dupont2024transcad,
  title={Transcad: A hierarchical transformer for cad sequence inference from point clouds},
  author={Dupont, Elona and Cherenkova, Kseniya and Mallis, Dimitrios and Gusev, Gleb and Kacem, Anis and Aouada, Djamila},
  booktitle={European Conference on Computer Vision},
  pages={19--36},
  year={2024},
  organization={Springer}
}

@article{elistratov2026cadevolve,
  title={CADEvolve: Creating Realistic CAD via Program Evolution},
  author={Elistratov, Maksim and Barannikov, Marina and Ivanov, Gregory and Khrulkov, Valentin and Konushin, Anton and Kuznetsov, Andrey and Zhemchuzhnikov, Dmitrii},
  journal={arXiv preprint arXiv:2602.16317},
  year={2026}
}

@article{fan2026caddesigner,
  title={CADDesigner: Conceptual CAD model generation with a general-purpose agent},
  author={Fan, Fengxiao and Ni, Jingzhe and Yin, Xiaolong and Wang, Sirui and Lu, Xingyu and Zou, Qiang and Tong, Ruofeng and Tang, Min and Du, Peng},
  journal={Computer-Aided Design},
  pages={104087},
  year={2026},
  publisher={Elsevier}
}

@book{foley1996computer,
  title={Computer graphics: principles and practice},
  author={Foley, James D},
  volume={12110},
  year={1996},
  publisher={Addison-Wesley Professional}
}

@article{kabisov2026cadreasoner,
  title={CADReasoner: Iterative Program Editing for CAD Reverse Engineering},
  author={Kabisov, Soslan and Kirichuk, Vsevolod and Volkov, Andrey and Savrasov, Gennadii and Barannikov, Marina and Konushin, Anton and Kuznetsov, Andrey and Zhemchuzhnikov, Dmitrii},
  journal={arXiv preprint arXiv:2603.29847},
  year={2026}
}

@inproceedings{khan2024cad,
  title={Cad-signet: Cad language inference from point clouds using layer-wise sketch instance guided attention},
  author={Khan, Mohammad Sadil and Dupont, Elona and Ali, Sk Aziz and Cherenkova, Kseniya and Kacem, Anis and Aouada, Djamila},
  booktitle={Proceedings of the IEEE/CVF Conference on Computer Vision and Pattern Recognition},
  pages={4713--4722},
  year={2024}
}

@article{khan2024text2cad,
  title={Text2cad: Generating sequential cad designs from beginner-to-expert level text prompts},
  author={Khan, Mohammad S and Sinha, Sankalp and Sheikh, Talha U and Stricker, Didier and Ali, Sk A and Afzal, Muhammad Z},
  journal={Advances in Neural Information Processing Systems},
  volume={37},
  pages={7552--7579},
  year={2024}
}

@inproceedings{kolodiazhnyi2025cadrille,
  title={cadrille: Multi-modal CAD Reconstruction with Reinforcement Learning},
  author={Kolodiazhnyi, Maksim and Tarasov, Denis and Zhemchuzhnikov, Dmitrii and Nikulin, Alexander and Zisman, Ilya and Vorontsova, Anna and Konushin, Anton and Kurenkov, Vladislav and Rukhovich, Danila},
  booktitle={The Fourteenth International Conference on Learning Representations},
  year={2025}
}

@article{kuhn1955hungarian,
  title={The Hungarian method for the assignment problem},
  author={Kuhn, Harold W},
  journal={Naval research logistics quarterly},
  volume={2},
  number={1-2},
  pages={83--97},
  year={1955},
  publisher={Wiley Online Library}
}

@inproceedings{li2026recad,
  title={Recad: Reinforcement learning enhanced parametric cad model generation with vision-language models},
  author={Li, Jiahao and Luo, Yusheng and Lou, Yunzhong and Zhou, Xiangdong},
  booktitle={Proceedings of the AAAI Conference on Artificial Intelligence},
  volume={40},
  number={8},
  pages={6190--6198},
  year={2026}
}

@inproceedings{li2023secad,
  title={Secad-net: Self-supervised cad reconstruction by learning sketch-extrude operations},
  author={Li, Pu and Guo, Jianwei and Zhang, Xiaopeng and Yan, Dong-Ming},
  booktitle={Proceedings of the IEEE/CVF Conference on Computer Vision and Pattern Recognition},
  pages={16816--16826},
  year={2023}
}

@inproceedings{li2026seek,
  title={Seek-CAD: A self-refined generative modeling for 3D parametric CAD using local inference via DeepSeek},
  author={Li, Xueyang and Li, Jiahao and Song, Yu and Lou, Yunzhong and Zhou, Xiangdong},
  booktitle={International Conference on Learning Representations},
  volume={2026},
  pages={94278--94298},
  year={2026}
}

@incollection{li2011globfit,
  title={Globfit: Consistently fitting primitives by discovering global relations},
  author={Li, Yangyan and Wu, Xiaokun and Chrysathou, Yiorgos and Sharf, Andrei and Cohen-Or, Daniel and Mitra, Niloy J},
  booktitle={Acm siggraph 2011 papers},
  pages={1--12},
  year={2011}
}

@inproceedings{liao2025automated,
  title={Automated cad modeling sequence generation from text descriptions via transformer-based large language models},
  author={Liao, Jianxing and Xu, Junyan and Sun, Yatao and Tang, Maowen and He, Sicheng and Liao, Jingxian and Yu, Shui and Li, Yun and Guan, Xiaohong},
  booktitle={Proceedings of the 63rd Annual Meeting of the Association for Computational Linguistics (Volume 1: Long Papers)},
  pages={21720--21748},
  year={2025}
}

@inproceedings{liu2024point2cad,
  title={Point2cad: Reverse engineering cad models from 3d point clouds},
  author={Liu, Yujia and Obukhov, Anton and Wegner, Jan Dirk and Schindler, Konrad},
  booktitle={Proceedings of the IEEE/CVF conference on computer vision and pattern recognition},
  pages={3763--3772},
  year={2024}
}

@inproceedings{ma2024draw,
  title={Draw step by step: Reconstructing cad construction sequences from point clouds via multimodal diffusion},
  author={Ma, Weijian and Chen, Shuaiqi and Lou, Yunzhong and Li, Xueyang and Zhou, Xiangdong},
  booktitle={2024 IEEE/CVF Conference on Computer Vision and Pattern Recognition (CVPR)},
  pages={27144--27153},
  year={2024},
  organization={IEEE}
}

@inproceedings{ma2023multicad,
  title={Multicad: Contrastive representation learning for multi-modal 3d computer-aided design models},
  author={Ma, Weijian and Xu, Minyang and Li, Xueyang and Zhou, Xiangdong},
  booktitle={Proceedings of the 32nd ACM International Conference on Information and Knowledge Management},
  pages={1766--1776},
  year={2023}
}

@inproceedings{oesau2016planar,
  title={Planar shape detection and regularization in tandem},
  author={Oesau, Sven and Lafarge, Florent and Alliez, Pierre},
  booktitle={Computer Graphics Forum},
  volume={35},
  number={1},
  pages={203--215},
  year={2016},
  organization={Wiley Online Library}
}

@inproceedings{qi2017pointnet,
  title={Pointnet: Deep learning on point sets for 3d classification and segmentation},
  author={Qi, Charles R and Su, Hao and Mo, Kaichun and Guibas, Leonidas J},
  booktitle={Proceedings of the IEEE conference on computer vision and pattern recognition},
  pages={652--660},
  year={2017}
}

@article{qi2026pointer,
  title={Pointer-cad: Unifying b-rep and command sequences via pointer-based edges \& faces selection},
  author={Qi, Dacheng and Wang, Chenyu and Xu, Jingwei and Chu, Tianzhe and Zhao, Zibo and Liu, Wen and Ding, Wenrui and Ma, Yi and Gao, Shenghua},
  journal={arXiv preprint arXiv:2603.04337},
  year={2026}
}

@inproceedings{rukhovich2025cad,
  title={Cad-recode: Reverse engineering cad code from point clouds},
  author={Rukhovich, Danila and Dupont, Elona and Mallis, Dimitrios and Cherenkova, Kseniya and Kacem, Anis and Aouada, Djamila},
  booktitle={Proceedings of the IEEE/CVF International Conference on Computer Vision},
  pages={9801--9811},
  year={2025}
}

@article{qi2026pointerv2,
  title={Pointer-CAD v2: Plan-Then-Construct CAD Generation with Dimension-Aware Parametric Precision},
  author={Qi, Dacheng and Wang, Chenyu and Xu, Jingwei and Ma, Yi and Gao, Shenghua},
  journal={arXiv preprint arXiv:2606.29301},
  year={2026}
}

@inproceedings{uy2022point2cyl,
  title={Point2cyl: Reverse engineering 3d objects from point clouds to extrusion cylinders},
  author={Uy, Mikaela Angelina and Chang, Yen-Yu and Sung, Minhyuk and Goel, Purvi and Lambourne, Joseph G and Birdal, Tolga and Guibas, Leonidas J},
  booktitle={Proceedings of the IEEE/CVF Conference on Computer Vision and Pattern Recognition},
  pages={11850--11860},
  year={2022}
}

@article{wang2024qwen2,
  title={Qwen2-vl: Enhancing vision-language model's perception of the world at any resolution},
  author={Wang, Peng and Bai, Shuai and Tan, Sinan and Wang, Shijie and Fan, Zhihao and Bai, Jinze and Chen, Keqin and Liu, Xuejing and Wang, Jialin and Ge, Wenbin and others},
  journal={arXiv preprint arXiv:2409.12191},
  year={2024}
}

@article{wang2020pie,
  title={Pie-net: Parametric inference of point cloud edges},
  author={Wang, Xiaogang and Xu, Yuelang and Xu, Kai and Tagliasacchi, Andrea and Zhou, Bin and Mahdavi-Amiri, Ali and Zhang, Hao},
  journal={Advances in neural information processing systems},
  volume={33},
  pages={20167--20178},
  year={2020}
}

@article{willis2021fusion,
  title={Fusion 360 gallery: A dataset and environment for programmatic cad construction from human design sequences},
  author={Willis, Karl DD and Pu, Yewen and Luo, Jieliang and Chu, Hang and Du, Tao and Lambourne, Joseph G and Solar-Lezama, Armando and Matusik, Wojciech},
  journal={ACM Transactions on Graphics (TOG)},
  volume={40},
  number={4},
  pages={1--24},
  year={2021},
  publisher={ACM New York, NY, USA}
}

@inproceedings{wu2021deepcad,
  title={Deepcad: A deep generative network for computer-aided design models},
  author={Wu, Rundi and Xiao, Chang and Zheng, Changxi},
  booktitle={Proceedings of the IEEE/CVF international conference on computer vision},
  pages={6772--6782},
  year={2021}
}

@inproceedings{wu2024point,
  title={Point transformer v3: Simpler faster stronger},
  author={Wu, Xiaoyang and Jiang, Li and Wang, Peng-Shuai and Liu, Zhijian and Liu, Xihui and Qiao, Yu and Ouyang, Wanli and He, Tong and Zhao, Hengshuang},
  booktitle={Proceedings of the IEEE/CVF conference on computer vision and pattern recognition},
  pages={4840--4851},
  year={2024}
}

@article{xu2024cad,
  title={Cad-mllm: Unifying multimodality-conditioned cad generation with mllm},
  author={Xu, Jingwei and Wang, Chenyu and Zhao, Zibo and Liu, Wen and Ma, Yi and Gao, Shenghua},
  journal={arXiv preprint arXiv:2411.04954},
  year={2024}
}

@inproceedings{xu2024pointllm,
  title={Pointllm: Empowering large language models to understand point clouds},
  author={Xu, Runsen and Wang, Xiaolong and Wang, Tai and Chen, Yilun and Pang, Jiangmiao and Lin, Dahua},
  booktitle={European Conference on Computer Vision},
  pages={131--147},
  year={2024},
  organization={Springer}
}

@article{xu2023hierarchical,
  title={Hierarchical neural coding for controllable cad model generation},
  author={Xu, Xiang and Jayaraman, Pradeep Kumar and Lambourne, Joseph G and Willis, Karl DD and Furukawa, Yasutaka},
  journal={arXiv preprint arXiv:2307.00149},
  year={2023}
}

@inproceedings{zhao2021point,
  title={Point transformer},
  author={Zhao, Hengshuang and Jiang, Li and Jia, Jiaya and Torr, Philip HS and Koltun, Vladlen},
  booktitle={Proceedings of the IEEE/CVF international conference on computer vision},
  pages={16259--16268},
  year={2021}
}

@inproceedings{zhu2023nerve,
  title={Nerve: Neural volumetric edges for parametric curve extraction from point cloud},
  author={Zhu, Xiangyu and Du, Dong and Chen, Weikai and Zhao, Zhiyou and Nie, Yinyu and Han, Xiaoguang},
  booktitle={Proceedings of the IEEE/CVF Conference on Computer Vision and Pattern Recognition},
  pages={13601--13610},
  year={2023}
}

@article{shao2024deepseekmath,
  title={Deepseekmath: Pushing the limits of mathematical reasoning in open language models},
  author={Shao, Zhihong and Wang, Peiyi and Zhu, Qihao and Xu, Runxin and Song, Junxiao and Bi, Xiao and Zhang, Haowei and Zhang, Mingchuan and Li, YK and Wu, Yang and others},
  journal={arXiv preprint arXiv:2402.03300},
  year={2024}
}

@inproceedings{you2025img2cad,
  title={Img2cad: Reverse engineering 3d cad models from images through vlm-assisted conditional factorization},
  author={You, Yang and Uy, Mikaela Angelina and Han, Jiaqi and Thomas, Rahul and Zhang, Haotong and Du, Yi and Chen, Hansheng and Engelmann, Francis and You, Suya and Guibas, Leonidas},
  booktitle={Proceedings of the SIGGRAPH Asia 2025 Conference Papers},
  pages={1--12},
  year={2025}
}

\end{document}